\documentclass[pdflatex,sn-aps]{sn-jnl}

\usepackage{graphicx}%
\usepackage{multirow}%
\usepackage{amsmath,amssymb,amsfonts}%
\usepackage{amsthm}%
\usepackage{mathrsfs}%
\usepackage[title]{appendix}%
\usepackage{xcolor}%
\usepackage{textcomp}%
\usepackage{manyfoot}%
\usepackage{booktabs}%
\usepackage{algorithm}%
\usepackage{algorithmicx}%
\usepackage{algpseudocode}%
\usepackage{listings}%

\usepackage{subcaption}
\usepackage{lineno}
\usepackage{lscape}

\newcommand{\figwdmat}{0.65in}          
\newcommand{\colwdmat}{0.65in}          

\theoremstyle{thmstyleone}%
\theoremstyle{thmstyletwo}%

\theoremstyle{thmstylethree}%

\begin{document}

\title[]{Sterol-induced raft-like domains in a model lipid monolayer}


\author*[1]{\spfx{S.} \fnm{Siva Nasarayya Chari} }\email{snchari@allduniv.ac.in}

\author[2]{\fnm{Bharat} \sur{Kumar}}\email{bharat@cuk.ac.in}
\equalcont{These authors contributed equally to this work.}

\affil*[1]{\orgdiv{Department of Physics, Faculty of Science}, \orgname{University of Allahabad}, \city{Prayagraj}, \postcode{211002}, \state{Uttar Pradesh}, \country{India}}

\affil[2]{\orgdiv{Department of Physics, School of Physical Sciences}, \orgname{Central University of Karnataka}, \city{Kalaburagi}, \postcode{585367}, \state{Karnataka}, \country{India}}


\abstract{A two-dimensional system consisting a mixture of highly coarse-grained saturated (S-type), unsaturated (U-type) lipid molecules, and cholesterol (C-type) molecules is considered to form a model lipid monolayer. All the S-, U- and C-type particles are spherical in shape, with distinct interaction strengths. The phase behavior of the system is studied for various compositions ($x$) of the C-type particles, ranging from $x = 0.1$ to $0.9$. The results show that a structurally ordered complex is formed with the S- and C-types in the fluid-like environment of U-type particles, for $x \in \lbrace 0.5 - 0.6\rbrace$. The time-averaged hexatic order parameter $\left\langle \Psi_{6} \right\rangle$ indicates that the dynamical segregation of S- and C-types exhibits a positional order, that is found to be maximum for $x$ in the range of 0.5 - 0.6. The mean change in the free energy ($\Delta G(x)$) obtained from the mean change in enthalpy ($\Delta H$) and entropy ($\Delta S$) calculations suggests that $\Delta G$ is minimum for $x \sim 0.6$. A phenomenological expression for the Gibbs free energy is formulated by explicitly accounting for the individual free energies of S-,U- and C-type particles and the mutual interactions between them. Minimizing this phenomenological $G$ with respect to the C-type composition results in the optimal value, $x^* = 0.564 \pm 0.001$ for stable coexistence of phases; consistent with the simulation results and also the previous experimental observations \cite{raghavendra_effect_2023}. All these observations signify the optimal C-type composition, $x \sim 0.5 - 0.6$.}




\maketitle

\section{Introduction} 
\label{sec:intro}

\noindent
The cellular processes of signal transduction, intercellular transport, and many fundamental mechanisms of animal cells are intricately connected with the lateral organization of lipids within the monolayer\cite{brown_functions_1998, sezgin_mystery_2017, goldfine_bacterial_1984}. The dynamical segregation of lipids facilitates the interaction of the cell membrane with proteins and other relevant functional residues. A membrane monolayer typically consists of saturated and unsaturated lipid molecules. Until the twentieth century, the membrane monolayer was perceived as a homogeneous collection of its constituents, as described by the fluid mosaic model\cite{singer_fluid_1972,nicolson_fifty_2022}. In the early twentieth century, it was understood that lipids in the cell membrane would indeed undergo a dynamical compartmentalization into small-scale domains \cite{nicolson_update_2013,sezgin_mystery_2017,lingwood_lipid_2010, munro_lipid_2003,delmas_importance_2013,pike_rafts_2006,nicolson_fluid_2023_book,nicolson_fluid_2023, heimburg_excitable_2023}, which are rich in cholesterol, sphingolipids, and other existing proteins to facilitate various inter- and intracellular functions. These domains are structurally ordered and, therefore, are denoted by the liquid-ordered ‘$L_o$’ phase. They often coexist with a fluid-like disordered environment denoted by the liquid-disordered phase '$L_d$’ \cite{risselada_molecular_2008,gomez_actively_2008,
simons_model_2004,rajendran_lipid_2005,lentz_ordered_1983,bhattacharya_interactions_2000} in their outer leaflet. These structurally ordered microdomains are technically termed as `lipid rafts' \cite{delmas_importance_2013,edidin_the_2003,sonnino_lipids_2010}, as originally hypothesized by Simons and Van Meer \cite{simons_lipid_1988} and later formally developed by Simons and Ikonen \cite{simons_functional_1997}. 
Lipid rafts play an important role in cell signaling \cite{cheng_biological_2019,pike_lipid_2003}, trafficking \cite{munro_lipid_2003,brown_functions_1998}, and signal transduction. To obtain a better understanding of the formation, structure, and dynamics of these domains, several experimental, theoretical, and computational approaches \cite{sviridov_lipid_2020,edidin_the_2003,mukai_lipid_2017,
holl_cell_2008,fan_formation_2010,mouritsen_lipid_2015,rog_cholesterol_2014,tieleman_computer_1997,
smondyrev_structure_1999,sarkar_minimal_2021} have recently been made, and possible applications in terms of therapeutic targeting \cite{sviridov_lipid_2020} have been investigated in invasive diseases such as cancer, T cell activation \cite{luo_the_2008} in HIV, etc. 

Goldfine et al. \cite{goldfine_bacterial_1984} studied the stability of bacterial membrane bilayer mixed with cholesterol and identified that excess of amount of unsaturated fatty acids, cholesterol and temperature significantly affect the stability of the bilayers. Kessel et al. \cite{kessel_interactions_2001} investigated the free energy difference related to the insertion of a single cholesterol molecule into a model lipid bilayer treating the insertion depth and the orientation as parameters. Rog et al. \cite{rog_cholesterol_2014} reviewed the effect of cholesterol and raft formation, based on the atomistic and coarse-grained model simulations. Pandit et al. \cite{pandit_complexation_2004} performed the Molecular Dynamics(MD) simulations of the DPPC and the DLPC bilayers mixed with 40 mol \% of cholesterol, and visualized the complexation. Tu et al. \cite{tu_constant-pressure_1998} studied the molecular dynmaics simulations of DPPC bilayer and observed no significant effect of 12.5 mol\% cholesterol on the bilayer conformations. Doi et al. \cite{doi_dpd_2024} carried out the Dissipative Particle Dynmaics (DPD) simulations of the LLC mixture and identified that the raft-like ordered microdomains can be replicated in LLC mixtures. Javanainen et al. \cite{javanainen_nanoscale_2017} performed the atomistic simulations of the saturated DPPC and cholesterol mixtures, which is a minimal standard for domain formation that explain a plethora of experimental observations. Smondyrev and Berkowitz \cite{smondyrev_structure_1999} used the MD simulations to study the DPPC bilayers mixed with low and high concentrations of cholesterol, and observed a reduction in the average area of membrane, and the total Gauche defects. Gu et al. \cite{gu_phase_2020} performed the atomistic simulations of DPPC:DOPC (binary), and DPPC:DOPC:Cholesterol (ternary) mixtures of lipid bilayers, and identified the compositions of 0.53:0.13:0.34 at temperature 280 K, for the system to be in $L_o$ phase. Sarkar and Farego \cite{sarkar_lattice_2023} carried out the Monte Carlo simulations of a DPPC/DOPC/Cholesterol ternary mixture using a lattice model for tunable domain sizes. They observed the respective composition of 0.46:0.16:0.38 for the $L_o$ phase at about 280 K temperature. Risselada and Marrink \cite{risselada_molecular_2008} conducted the model membrane simulations of ternary mixtures of saturated and unsaturated lipids along with cholesterol and noticed spontaneous separation into the structurally ordered $L_o$ and disordered $L_d$ phases. 

\par 
Hammond et al. \cite{hammond_crosslinking_2005} experimentally studied the effect of crosslinking ganglioside GM1 membrane components and observed the uniform membrane phase separate into a coexisting $L_o$ and $L_d$ phases. Bhattacharya and Haldar \cite{bhattacharya_interactions_2000} studied the bilayer and cholesterol mixture using the methods of steady-state flourescence anisotropy, X-ray diffraction of lipid-cholesterol coaggregated film, and 1H-NMR spectroscopy, and emphasized that the cholesterol induced effects on the lipid bilayer can not be fully understood on the basis of the Hydrogen bonds. Edidin \cite{edidin_state_2003} reviewed the state of the lipid rafts from model membrane computations to the experimental observations, discussing the scope for further investigations. Hao et al. \cite{hao_thermodynamic_2009} conducted the experiments of cholesterol mixed D-sphingosine using the Langmuir-Blodgett monolayers. AFM investigations revealed the attractive interactions between molecules in the presence of cholesterol. Ratajczak et al. \cite{ratajczak_ordered_2009} performed experiments with sphingomyelin-dihydrocholesterol (SM-DChol) monolayers. With the help of grazing incidence X-ray diffraction (GIXD), they identified that the $d-$spacing increases linearly after 35 mol\% of DChol, and this linear relationship holds until very high mole fractions of DChol. Crane and Tamm \cite{crane_role_2004} investigated the cholesterol-induced formation of the $L_o / L_d$ phases through fluorescence microscopy in ternary mixtures of porcine brain phosphatidylcholine (bPC), porcine brain sphingomyelin (bSM), and cholesterol. They observed that the fraction of the $L_o$ phase is governed by the cholesterol concentration, and there exists a percolation threshold of 40-50\% of cholesterol where network of $L_o$ starts interconnecting. 

Recent experimental observations in lipid monolayers mixed with cholesterol resulted in the formation of a condensed complex between cholesterol and saturated lipids\cite{rosenhouse-dantsker_cholesterol_2019,raghavendra_effect_2023,veatch_organization_2002,brown_functions_1998,reitveld_differential_1998}. However, unsaturated lipid molecules tend to induce ‘fluctuations’ in the orientational order that appears in the system \cite{kessel_interactions_2001,mukai_lipid_2017}, due to their relatively long alkyl chains. Hence, the complex formed is mainly between saturated lipids and cholesterol molecules. Furthermore, it was observed that the tendency to form such complexes in the system increased with an increase in the mole fraction of cholesterol to an optimal composition \cite{raghavendra_effect_2023}. 

Taking the above observations, here we propose a simple and highly coarse-grained model to describe the formation of ordered domains in a lipid monolayer in the presence of a given cholesterol composition. In the model, saturated, and unsaturated lipid molecules and cholesterol are considered to be spherical particles interacting via a Lennard-Jones-type potential with distinct interaction strengths. The dynamics of this ternary mixture of particles is investigated for emerging lateral organization. Particles are labelled S-type (for saturated), U-type (for unsaturated lipid molecules), and C-type (for cholesterol). Further numerical details of the model and the method of simulation are discussed in the next section.

\par
In summary, this work aims to observe the formation of ordered microdomains in a highly coarse-grained model of a lipid monolayer when a specific composition ($x$) of cholesterol is present. The goal is to find an optimal $x$ value at which the system achieves the maximum positional order in the microdomains and the minimum free energy. 

\section{Methodology}
\label{sec:methods}
\subsection{Description of the model}
\label{subsec:model}

Both S- and U-type particles are taken to be equal in size, $\sigma_S = \sigma_U = 1.0$, and the cholesterol (C-type) particles are considered to be relatively larger, $\sigma_C = 1.01$, in units of $\sigma_S$. Cholesterol in general will have a larger steroidal core than saturated or unsaturated lipid molecules. However, a higher disparity in particle size would result in entropic differences among the constituents. We have performed the simulations by systematically varying the difference in the size of the C-type particle from $\lbrace 0\%, 2\%, 5\%, \rm{and}~ 10\% \rbrace$, and obtained the behavior of enthalpy and entropy as a function of cholesterol composition $x$, see fig. SF1 in the Supplementary Information(SI). From fig. SF1, we observe that higher the size of the C-type, the greater the entropy that the system generates. Therefore, a $1\%$ difference in size is assumed for the C-type, without loss of generality. 

Pairwise interactions between the S and C types are considered Lennard-Jones 6-12 potentials that are truncated and shifted at $r = 2.5\sigma_S$ for smooth variation of the force. The remaining pairwise interactions are given by the Weeks-Chandler-Andersen (WCA) potential. Furthermore,
\begin{align}
	\sigma_{\alpha} &= 1.0, ~~\mathrm{and}~, \\
	\epsilon_{\alpha\alpha} &=1.0 ~~\mathrm{where} ~~\alpha \in \{ S, U \}~,
\end{align}

whereas the size and interaction parameters concerning the C-type particles are,
\begin{align}
	\sigma_C &= 1.01 ~, \\
	\epsilon_{CC} &= \epsilon_{UC} = 1.0~, ~~\mathrm{and} \\
	\epsilon_{SC} &= n_{C}/N~,
\end{align}

where, $n_C$ and $N$ denote the number of C-type particles and the total number of particles in the system respectively, 
\begin{equation}
	n_{S} + n_{U} + n_{C} = N~.
	\label{eq:conserveN}
\end{equation}

The Lorentz-Berthelot rules\cite{lorentz_ueber_1881, berthelot_sur_1898} are followed to mix the interactions between different types, except for the S-C pair, 
\begin{align}
	\sigma_{\alpha\beta} &= \frac{\sigma_{\alpha} + \sigma_{\beta}}{2} ~~\mathrm{for} ~~\alpha, ~\beta \in \{ S, U, C \}, \\
	\epsilon_{\alpha \beta} &= \sqrt{\epsilon_{\alpha\alpha} \epsilon_{\beta\beta}} ~~\mathrm{for} ~~\alpha, ~\beta \in \{ S, U, C \} ~~\mathrm{except ~for} \\
	\epsilon_{SC} &= \epsilon_{CS} = \frac{n_C}{N} ~.
\end{align}

It is to note here that the seminal contributions from the work of McConnell\cite{anderson_phase_2000, radhakrishnan_cholesterolphospholipid_1999} says that the interactions within the lipid monolayer in the air-water interface go beyond the pairwise in the presence of the cholesterol. However in this work, the interactions between the S-, U-, and C-type particles are considered pairwise, but the S-C interaction strength is made as a linear function of the cholesterol mole fraction. Though clearly this is not a 3-body interaction, it could be considered a mean field alternative of the same. 

\subsection{Numerical details}
\label{subsec:numerical}

We begin with an initial configuration with all the particles ($N=1000$) randomly distributed within the two-dimensional simulation box, as per the desired density $\rho^*$, and the mole fraction of the C-type particles,
\begin{equation}
	x = \frac{n_C}{N}~.
\end{equation}

The number of S- and U-type particles would then be, $n_U = n_S = (1-x)N/2$, satisfying eq.(\ref{eq:conserveN}). The system is then thermalized at a non-dimensionalized temperature $T^* = 0.1$ using a Nos\'{e}-Hoover thermostat\cite{nose_molecular_1984,nose_unified_1984,hoover_canonical_1985} with the temperature damping parameter value equal to 100 time units. The interparticle forces are initially obtained from a soft potential to avoid blow-up of energy due to possible overlaps in the initial random configuration, 
\begin{align}
	V(r) &= A\left[1 + \cos\left({\frac{\pi r}{r_c}}\right)\right] ~~\mathrm{for} ~~ r < r_c ~,
\end{align}

\noindent
where $r_{c} = 2^{1/6}\sigma$ is chosen as the cut-off distance. After proper thermalization the interaction parameters were redefined as described in subsection ~\ref{subsec:model},
\begin{align}
	V_{\alpha\beta}(r) =  V_{LJ}^{\alpha\beta}(r) - V_{LJ}^{\alpha\beta}(r=r_c) ~~~~\mathrm{for}~(r<r_c)~,\\
	\mathrm{where}~~ V_{LJ}^{\alpha\beta}(r) = 4\epsilon_{\alpha\beta}\left[ \left(\frac{\sigma_{\alpha\beta}}{r}\right)^{12} - \left(\frac{\sigma_{\alpha\beta}}{r}\right)^{6} \right]~,
\end{align} 
where $\alpha, \beta \in \lbrace S, C\rbrace$. Otherwise,
\begin{align}
	V_{U\beta}(r) = V_{LJ}^{U\beta}(r)+\epsilon_{U\beta}~~~\mathrm{for}~(r<2^{1/6}\sigma_{U\beta})~,
\end{align}
in the above, $\beta \in \lbrace S,C,U\rbrace$.
The system is then equilibrated with new potential parameters after coupling it to a Berendsen thermostat \cite{berendsen_molecular_1984} at temperature $T^* = 0.1$, with a coupling constant of $\tau = 10\delta t$. 
The Berendsen thermostat involves a velocity rescaling mechanism with the associated relaxation parameter $\tau$, whereas the Nos\'e-Hoover thermostat takes an extended Hamiltonian approach for the system $+$ reservoir. Therefore, the Berendsen thermostat relaxes the system to the target temperature faster than the Nos\'e-Hoover thermostat. We have also performed the simulations using the Nos\'e-Hoover thermostat, and the results of the same are compared with the results from the Berendsen thermostat, for $\rho^* = 0.3$, and C-type size $1.02 \sigma_S$. Please see fig. SF2, and SF3 in the SI, which shows the behavior of thermodynamic variables of the system with the C-type composition for different thermostats. 
In order to sample the configurations in the respective NVT ensemble, the system's equations of motion are time-integrated up to 2 ns, with $\delta t = 0.002$. A set of independent simulations are performed for various values of the C-type composition, ($x=0.1$ to $0.9$ in steps of $0.1$) and the density of the system $\rho^* = \{0.1, 0.3, 0.5, ~\mathrm{and} ~0.7\}$. All the simulations reported here are performed using the open-source molecular dynamics simulation package, LAMMPS\cite{noauthor_lammps_nodate}. Particles' equations of motion are time integrated using the Velocity-Verlet scheme with the periodic boundary conditions. 

\section{Results and Discussion}
\label{sec:results}

\begin{figure*}
	\begin{tabular}{l p{\colwdmat} p{\colwdmat} p{\colwdmat} p{\colwdmat} p{\colwdmat} p{\colwdmat} }
		~              & $x = 0.1$ & $x = 0.4$ & $x = 0.5$ & $x = 0.6$ & $x = 0.7$ & $x = 0.9$ \\
		\rotatebox{90}{~~~~$\rho^* = 0.1$} & \includegraphics[width=\figwdmat]{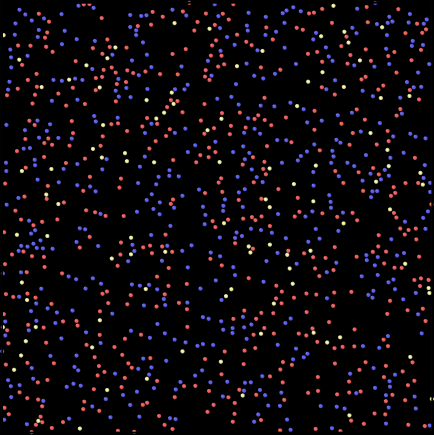} & \includegraphics[width=\figwdmat]{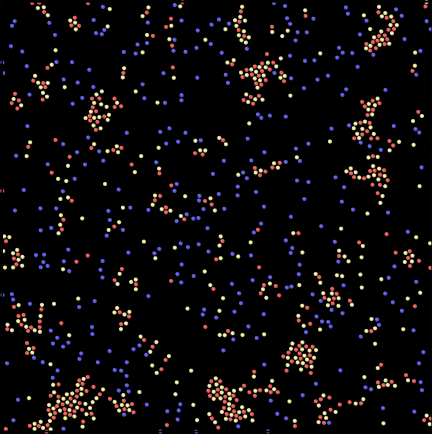} & \includegraphics[width=\figwdmat]{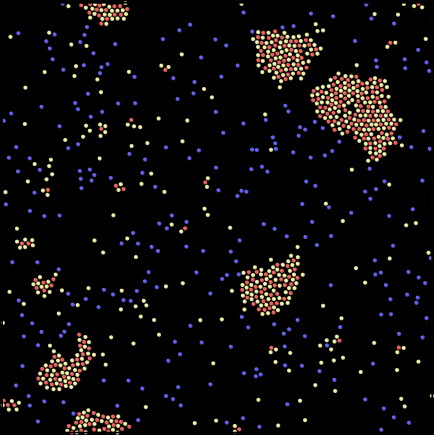} 
		& \includegraphics[width=\figwdmat]{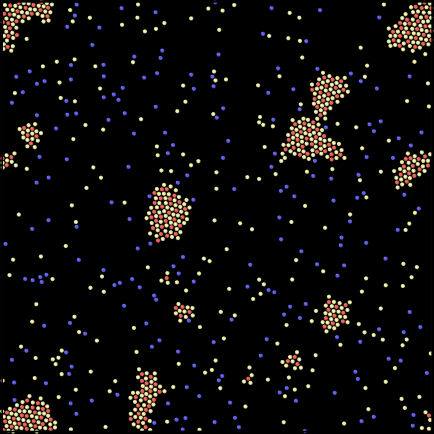}
		& \includegraphics[width=\figwdmat]{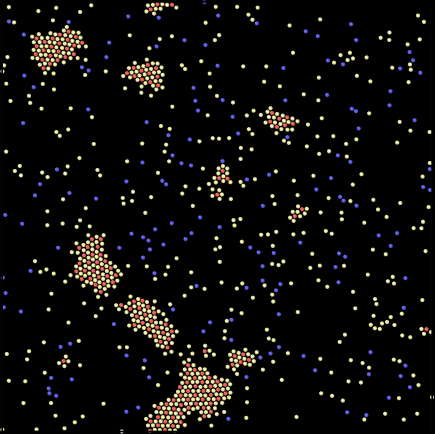}
		& \includegraphics[width=\figwdmat]{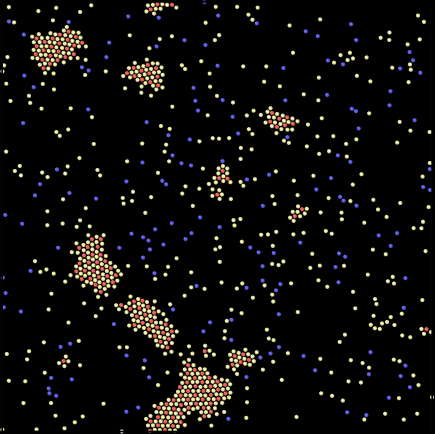} \\
		\rotatebox{90}{~~~~$\rho^* = 0.3$} & \includegraphics[width=\figwdmat]{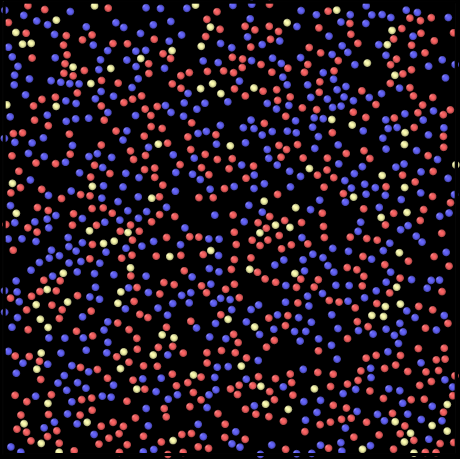} & \includegraphics[width=\figwdmat]{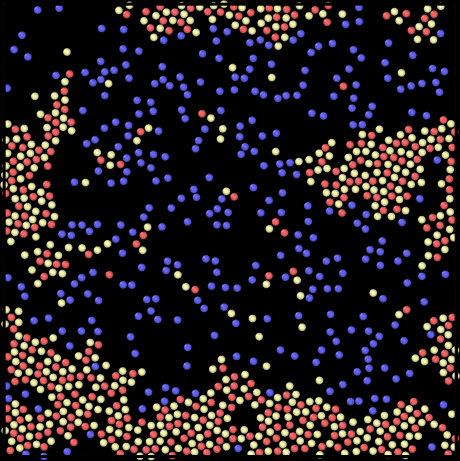} & \includegraphics[width=\figwdmat]{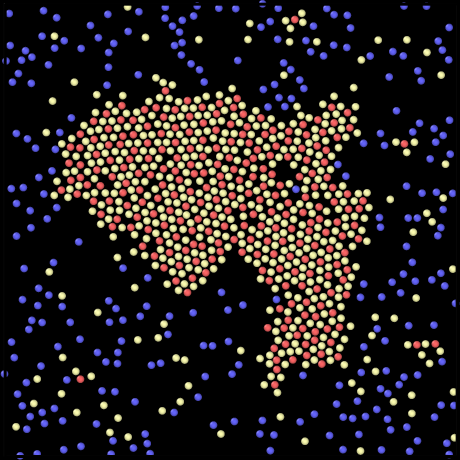} 
		& \includegraphics[width=\figwdmat]{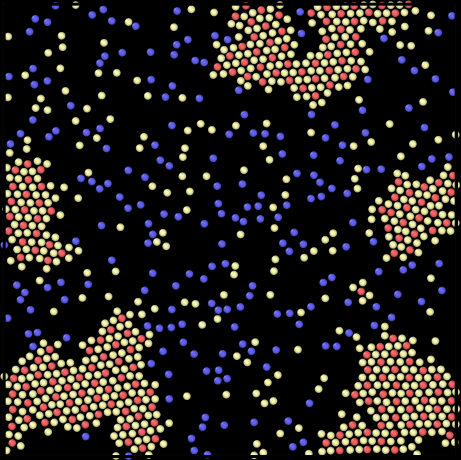}
		& \includegraphics[width=\figwdmat]{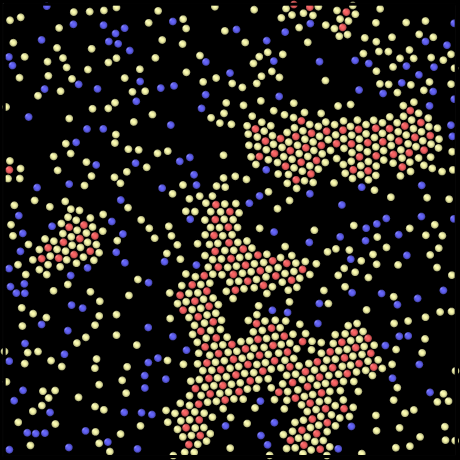}
		& \includegraphics[width=\figwdmat]{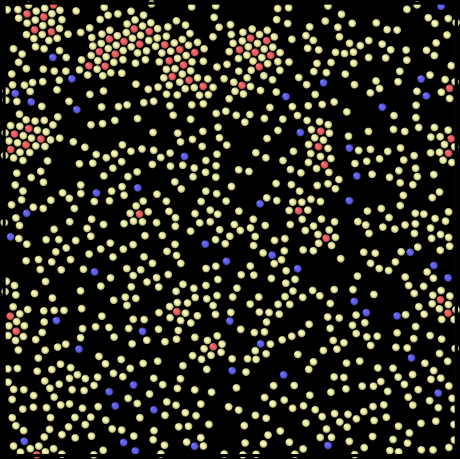} \\
		\rotatebox{90}{~~~~$\rho^* = 0.5$} & \includegraphics[width=\figwdmat]{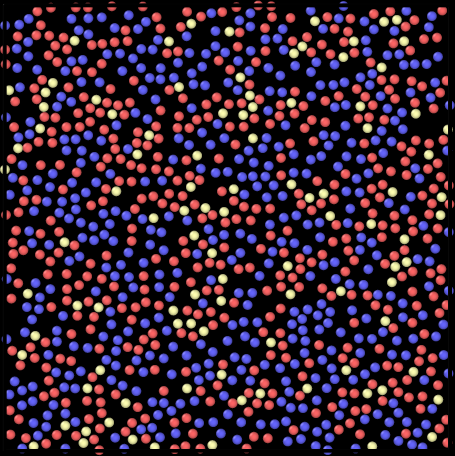} & \includegraphics[width=\figwdmat]{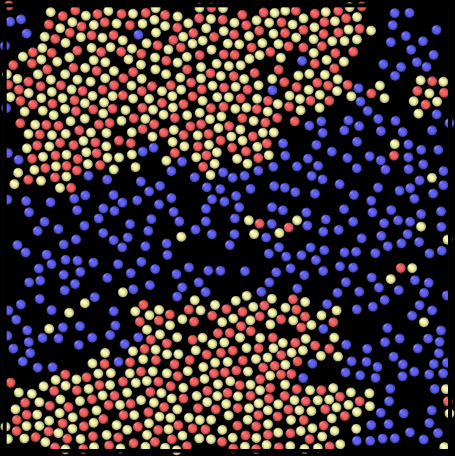} & \includegraphics[width=\figwdmat]{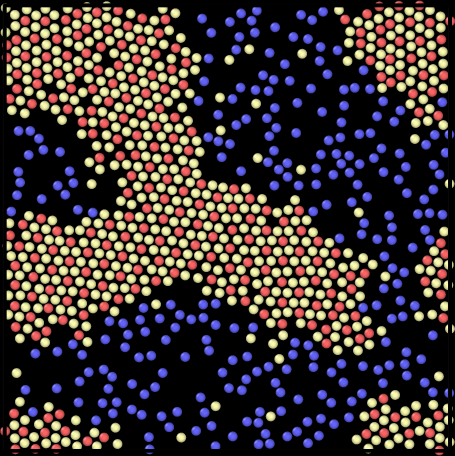} 
		& \includegraphics[width=\figwdmat]{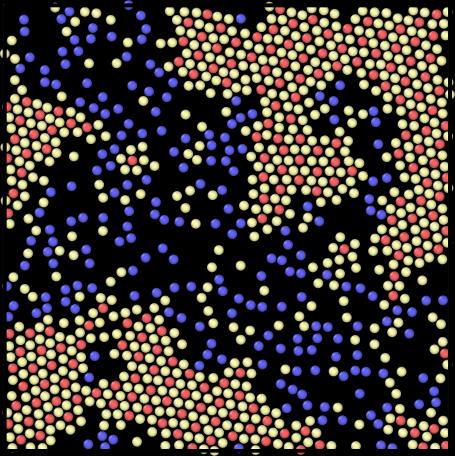}
		& \includegraphics[width=\figwdmat]{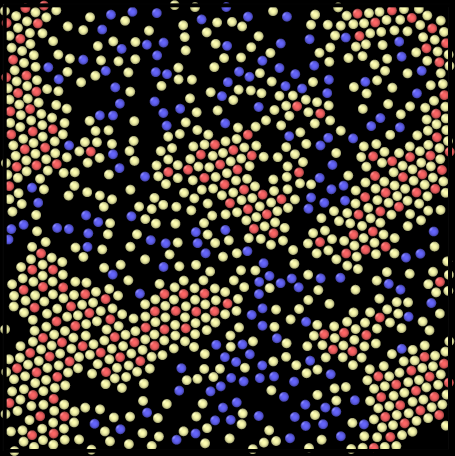}
		& \includegraphics[width=\figwdmat]{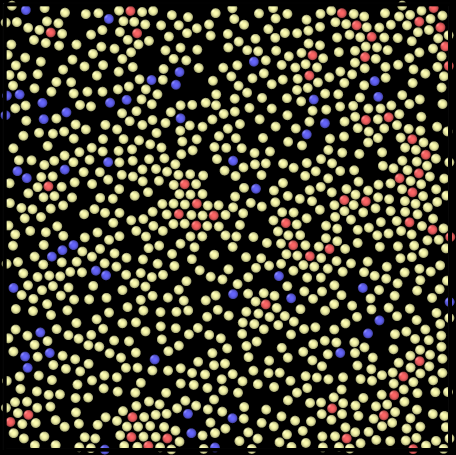} \\
		\rotatebox{90}{~~~~$\rho^* = 0.7$} & \includegraphics[width=\figwdmat]{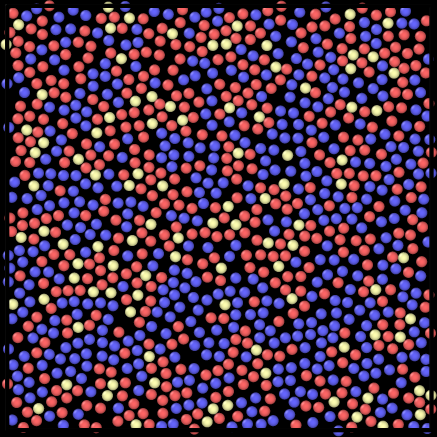} & \includegraphics[width=\figwdmat]{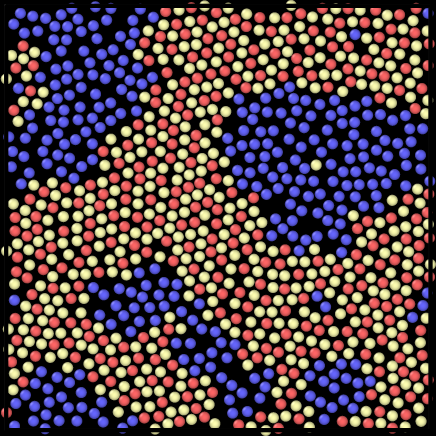} & \includegraphics[width=\figwdmat]{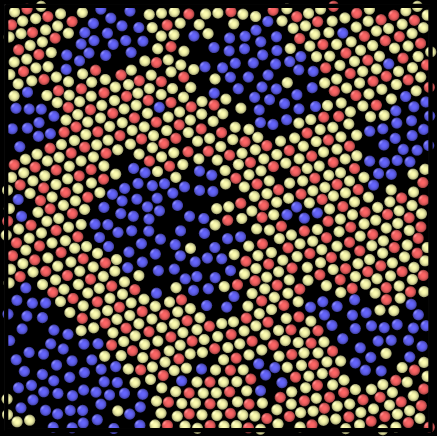} 
		& \includegraphics[width=\figwdmat]{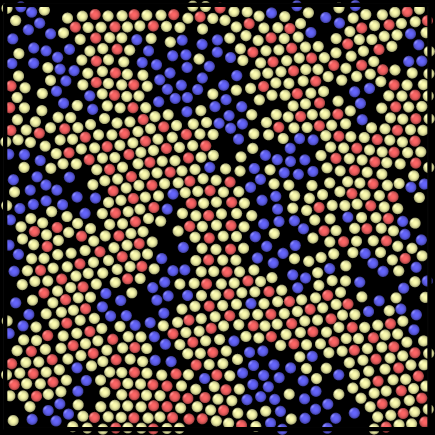}
		& \includegraphics[width=\figwdmat]{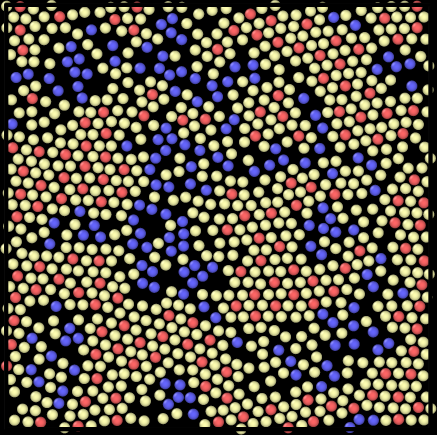}
		& \includegraphics[width=\figwdmat]{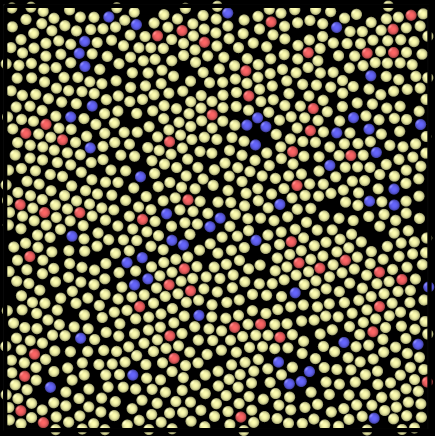} \\
	\end{tabular}
	\caption{(color online) Screenshots of the system at various compositions of C-type, \{$x = 0.1, ~0.4, ~0.5, ~0.6, ~0.7 ~~\mathrm{and} ~0.9$\}, at density \{$\rho^* = 0.1, ~0.3, ~0.5, ~~\mathrm{and} ~0.7$\}, and temperature $T^*=0.1$. Yellow, Red, and Blue spheres represent the C, S and U-type particles, respectively.}
	\label{fig:one}
\end{figure*}

\noindent
From fig. (\ref{fig:one}), we observe the formation of a complex or a microdomain between the S- and C-type particles. It is also noticed that for $0.5 \le x \le 0.6$, the complex acquires a structural order, which is an intercalation of a honeycomb lattice of C-type (yellow color) particles and a triangular lattice of S-type (red color) particles. In this range of $x$ values, most of the C-type is completely engaged by the S-type as $n_C \simeq 2 n_S$, satisfying the basic requirement for such an intercalated \textit{lattice} arrangement. Further increase in the $x$ increases the non-engaged C-type particles, resulting in an increase in entropy.

\begin{figure*}
	\centering
	\begin{subfigure}{0.48\linewidth}
		\includegraphics[width=0.99\textwidth]{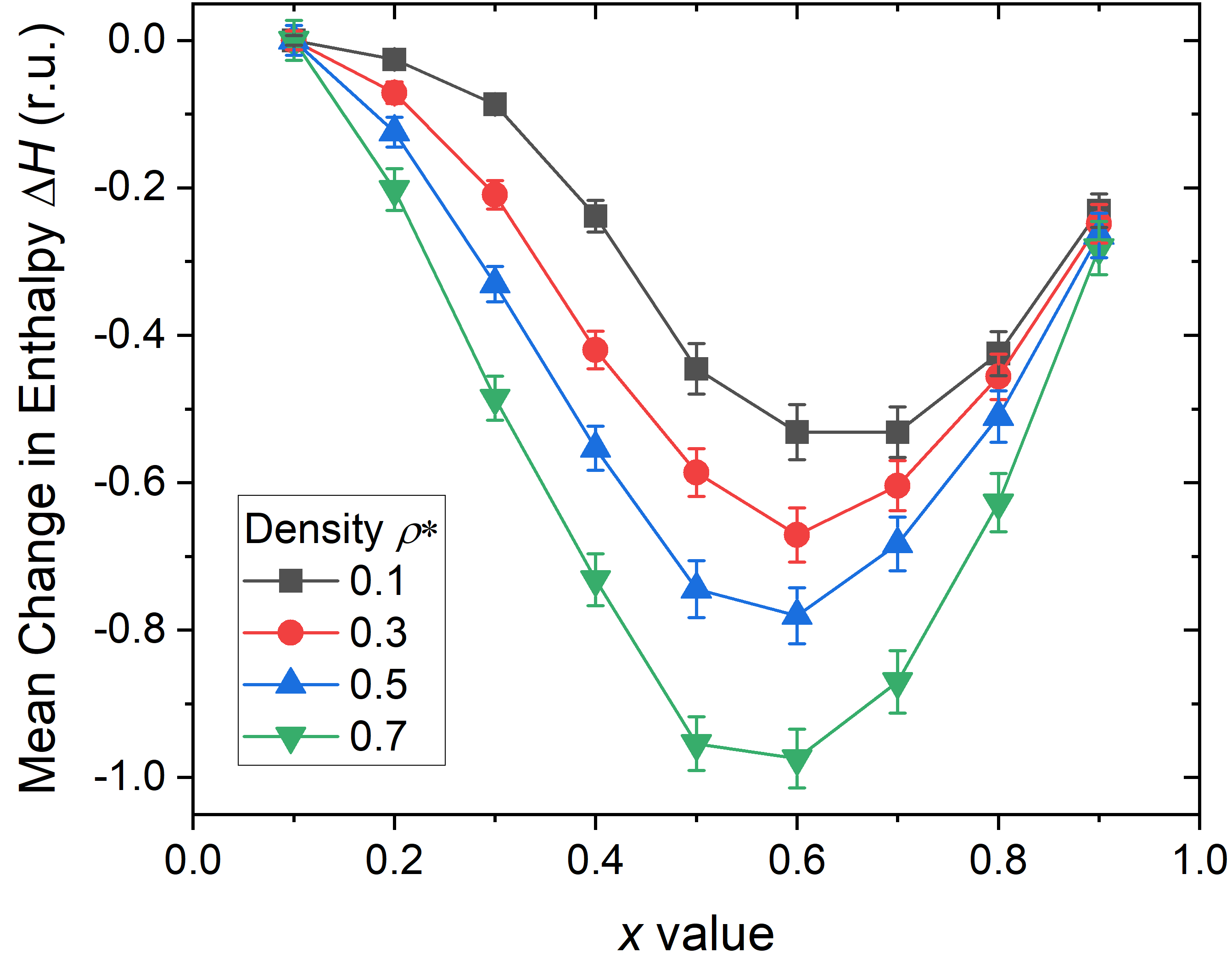}
		\caption{}
	\end{subfigure}
\hspace{3mm}
	\begin{subfigure}{0.48\linewidth}
		\includegraphics[width=0.99\textwidth]{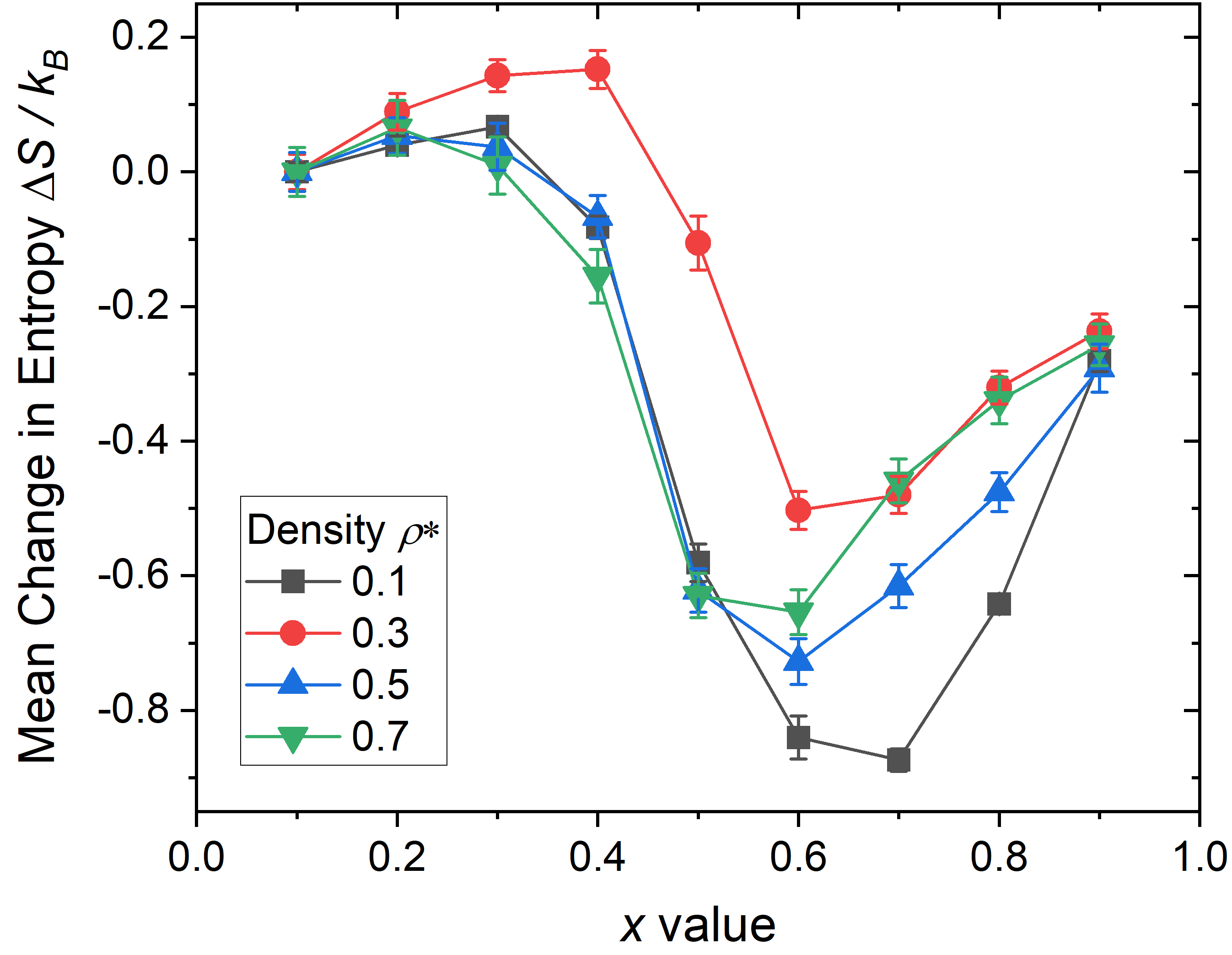}
		\caption{}
	\end{subfigure} \\[2mm]
	\begin{subfigure}{0.48\linewidth}
        \vspace{-4mm}
		\includegraphics[width=0.99\textwidth]{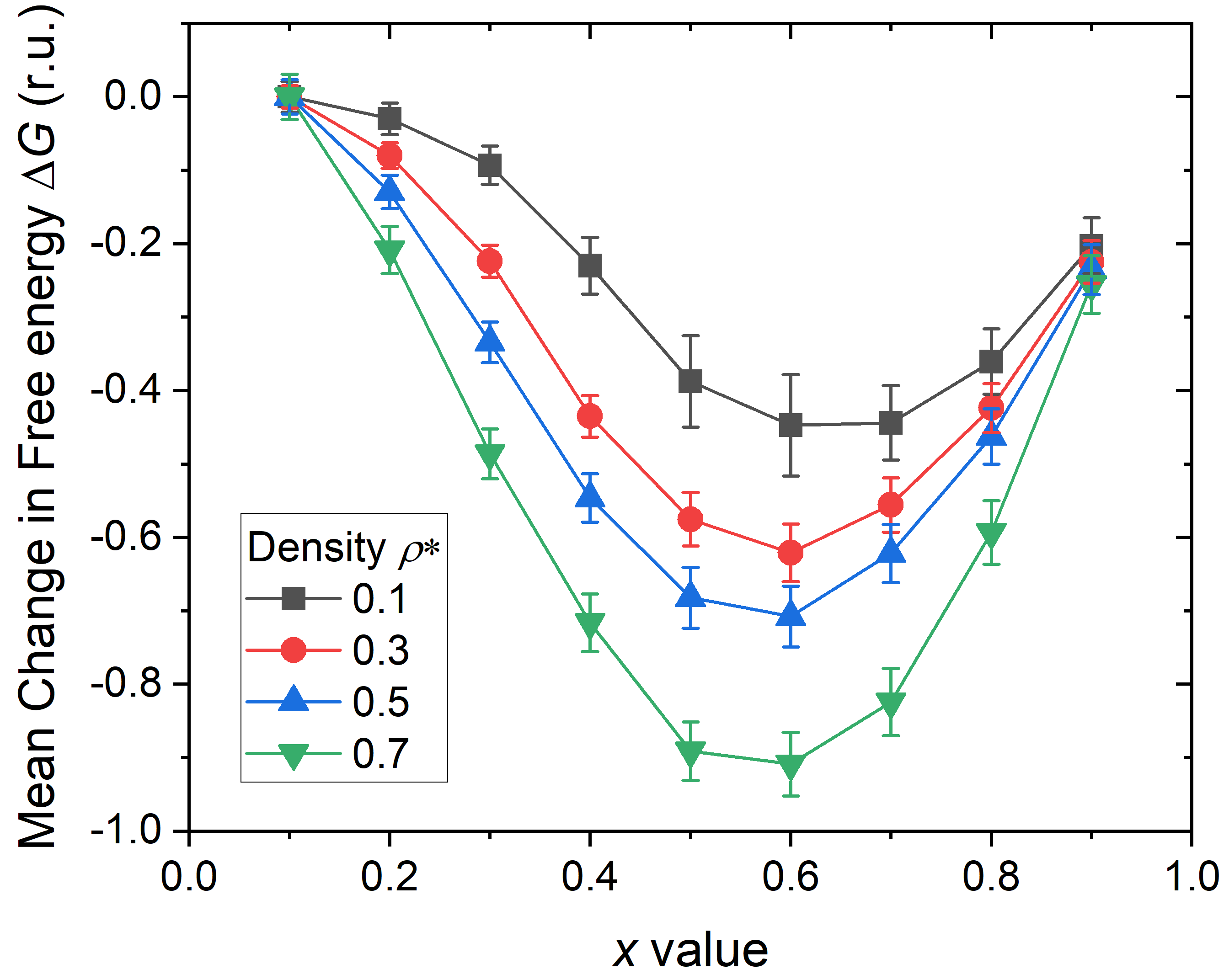}
		\caption{}
	\end{subfigure}
\hspace{3mm}
	\begin{subfigure}{0.48\linewidth}
        \vspace{-4mm}
		\includegraphics[width=1.0\textwidth]{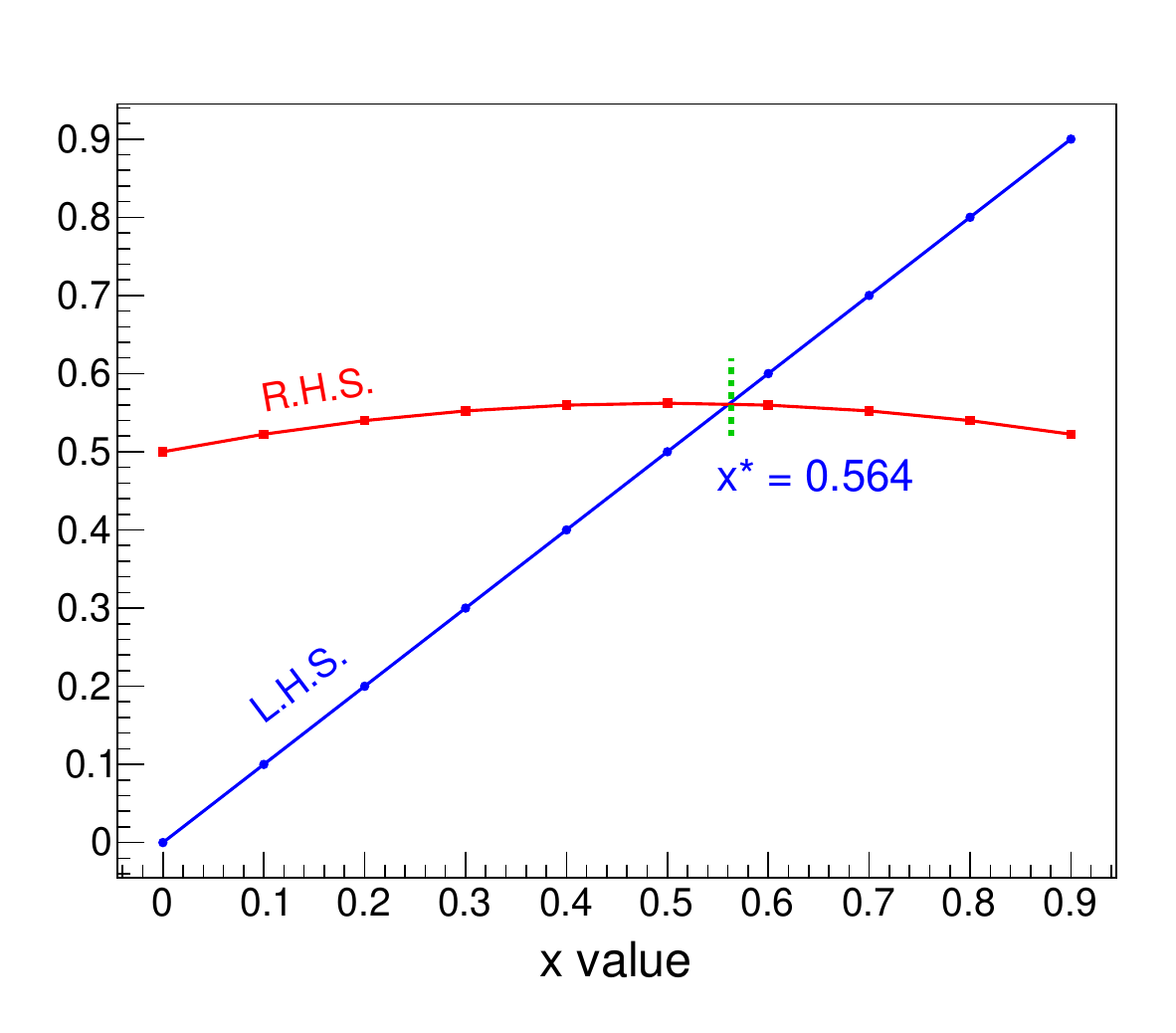}
		\caption{}
	\end{subfigure}
	
	\caption{(Color online) Change in (a). Enthalpy ($H = U + PV$), (b). Entropy and (c). Free energy averaged over several steady-state configurations, observed relative to their initial values. From the figure we observe that the $\Delta G$ is minimum around $x = 0.6$. (d). Graphical solution to the transcendental equation (\ref{eq:transcend}). }
	\label{fig:two}    
\end{figure*}

The mean change in the enthalpy ($\Delta H$) is calculated from the simulation for all the considered values of $\rho^*$ and $x$. The value of $\Delta H$ is observed with reference to its value at $x = 0.1$. i.e., ($\Delta H = \Delta H_{sys}(x) - \Delta H_{sys}(x = 0.1)$), see fig. (\ref{fig:two}) for more details. From fig. \ref{fig:two}(a) we observe that the change in enthalpy is minimum around $x = 0.6$, for almost all values of $\rho^*$. The pair entropy\cite{piaggi_entropy_2017,nettleton_expression_1958} is calculated for each coarse-grained particle,
\begin{equation}
	S_{i} = -2\pi\rho k_{B}\int_{0}^{r_m}{\left[g(r)\ln(g(r))-g(r) + 1\right] r^{2}dr}~, 
\end{equation}
where $r_m$ is the maximum distance up to which the neighbors are considered while evaluating the radial distribution function. Here, the ruggedness in the $g(r)$ is smoothed out using a Gaussian distribution,
\begin{equation}
	g^i(r) = \frac{1}{4\pi\rho r^2}\sum_{j=1}^{n_b}{\frac{1}{\sqrt{2\pi\xi^{2}}} e^{-(r-r_{ij})^2 / 2\xi^2}}~.
\end{equation}
In the above, the sum $j$ is over all the neighbors of $i$th particle, that exist within a distance of $r_m (= 2\sigma)$, and the parameter $\xi (= 0.125)$ is used as a convenient control parameter for smoothing. The parameter $S_i$ distinguishes the ordered arrangement of particles in a disordered fluid-like environment. Negative values of $S_i$ indicates the prevailing structural order in the locality of $i$th particle. More negative values represent the high extent of order in the same. The so calculated pair entropy of each particle is summed for the whole system and is averaged over a number of steady state configurations. In fig. \ref{fig:two}(b) we depicted the mean pair entropy of the system observed with reference to its value at $x = 0.1$, ($\Delta S = \Delta S_{sys}(x) - \Delta S_{sys}(x = 0.1)$) From the figure, we notice that there is more structural order around $x = 0.6$, for almost all the observed $\rho^*$ values. We determined the mean change in free energy from the above $\Delta H$, and $\Delta S$ values. From fig. \ref{fig:two}(c) we observe that the mean change in free energy is minimum around $x = 0.6$, at almost all values of $\rho^*$. From fig. (\ref{fig:two}) we understand that the system is more stable around $x = 0.6$.

\noindent

\subsection{Calculation from excess free energy}
\label{subsec:free_energy}

The composition-dependent free energy of a membrane-cholesterol system is highly dependent on the interactions between lipids and cholesterol, which can modulate the properties of the bilayer \cite{kessel_interactions_2001}. We have formulated a phenomenological expression for the Gibbs free energy of the system as a function of the cholesterol composition, $x$ as shown below. 

\begin{align}
    G &= x G_{C} +  \frac{(1-x)}{2} G_{S} + \frac{(1-x)}{2} G_{U} + U_{SC} \frac{x(1-x)}{2} + U_{SU} \frac{(1-x)^2}{4} + U_{CU} \frac{x(1-x)}{2} + \nonumber \\
	&+ Nk_{B}T\left[x\ln(x) + \frac{(1-x)}{2}\ln{\frac{(1-x)}{2}} + \frac{(1-x)}{2}\ln{\frac{(1-x)}{2}} \right]~,\nonumber \\
\implies	\tilde{G} &= \frac{G}{k_B T} = x \tilde{G}_{C} +  \frac{(1-x)}{2} \tilde{G}_{S} + \frac{(1-x)}{2} \tilde{G}_{U} + \tilde{U}_{SC} \frac{x(1-x)}{2} + \tilde{U}_{SU} \frac{(1-x)^2}{4} +  \nonumber \\
	&+ \tilde{U}_{CU} \frac{x(1-x)}{2} + N\left[x\ln(x) + (1-x)\ln(1-x) - (1-x)\ln(2)\right]~.
	\label{eq:dG_main}
\end{align}

In the above eq. (\ref{eq:dG_main}), the first three terms represent the the free energy of the individual C-, S-, and U-type particles, fourth, fifth, and sixth terms represent the interaction terms betwen the S-C, S-U, and C-U type pairs respectively. The last term in square brackets is due to the mixing free energy contribution to $G$. Following the non-dimensionalization of $G$, we define, 
\begin{align*}
    \tilde{U}_{\alpha \beta} = \frac{U_{\alpha \beta}}{k_B T}, ~\textrm{where} ~~\alpha, \beta \in \lbrace S, C, U\rbrace.
\end{align*}

Minimizing $\tilde{G}$ with respect to the C-type composition $x$ will lead to an optimal composition value $x^*$ at which $\tilde{G}$ is minimum, along the composition space of C-type. Though the compositions of the other two types could be treated independently, the formulated free energy do not encircle such possibility. Our aim here is to seek for the optimal cholesterol (C-type) composition, phenomenologically. Therefore, we demand 
\begin{align}
	  \frac{d(\tilde{G})}{dx} & = 0 \\	
	\implies \left[ \tilde{g} + \left(\frac{1}{2}-x\right) \tilde{u} - \frac{\tilde{U}_{SU}}{4} \right] &= -N\ln\left( \frac{2x}{1-x}\right)~ ,
	\label{eq:19}
\end{align}
where 
\begin{align}
    \tilde{g} = \tilde{G}_C - \left[ \frac{\tilde{G}_S + \tilde{G}_U}{2}\right]~, ~\textrm{and}~~ 
    \tilde{u} = \tilde{U}_{SC} + \tilde{U}_{CU} - \frac{\tilde{U}_{SU}}{2}~. 
    \label{eq:20}
\end{align}
This results in a transcendental equation in $x$,  
\begin{equation}
	x = \frac{1}{1 + 2 \exp\lbrace \left[ \tilde{g} + \left(\frac{1}{2} - x \right)\tilde{u} - \frac{\tilde{U}_{SU}}{4} \right]/N \rbrace}~.
	\label{eq:transcend}
\end{equation}
We require the parameter values of $\tilde{g}, \tilde{u}$, and  $\tilde{U}_{SU}$ in order to solve eq.(\ref{eq:transcend}) graphically, for $x^*$. The value of $\tilde{g}/N = -0.693$ as extracted from the individual chemical potential values of lipid molecules and cholesterol, mentioned in \cite{anderson_phase_2000}. The value of $\tilde{u}/N = \frac{1}{2}-x$ as evaluated from eq. (\ref{eq:20}) according to the interactions considered in the simulation, $\tilde{U}_{SC}/N = -x, \tilde{U}_{CU}/N = 1.0$ and $\tilde{U}_{SU}/N = 1.0$. 

Now, eq. (\ref{eq:transcend}) can be solved graphically by incorporating these parameter values; intersection point of the two curves (see fig. \ref{fig:two}(d)) at $x^* \sim 0.564$ becomes the solution of eq. (\ref{eq:transcend}). Hence the optimal composition $x^*$ obtained from the phenomenological free energy is consistent with the simulation results. 

\subsection{Positional order in the microdomains}

\begin{figure*}
	\centering
	\begin{subfigure}{0.48\textwidth}
		\includegraphics[width=0.99\textwidth]{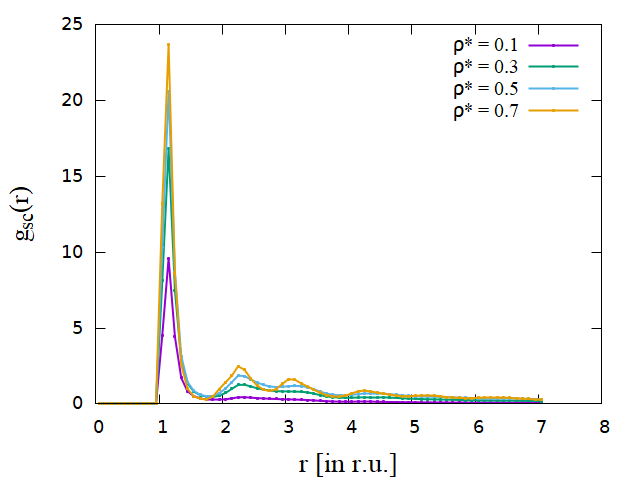}
		\caption{At $x = 0.4$}
	\end{subfigure}
	\hspace{2mm}
	\begin{subfigure}{0.48\textwidth}
		\includegraphics[width=0.99\textwidth]{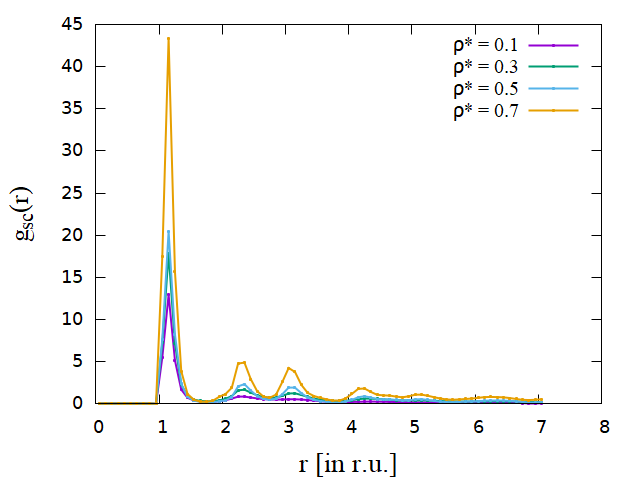}
		\caption{$x = 0.5$}
	\end{subfigure} \\
	\begin{subfigure}{0.48\textwidth}
		\includegraphics[width=0.99\textwidth]{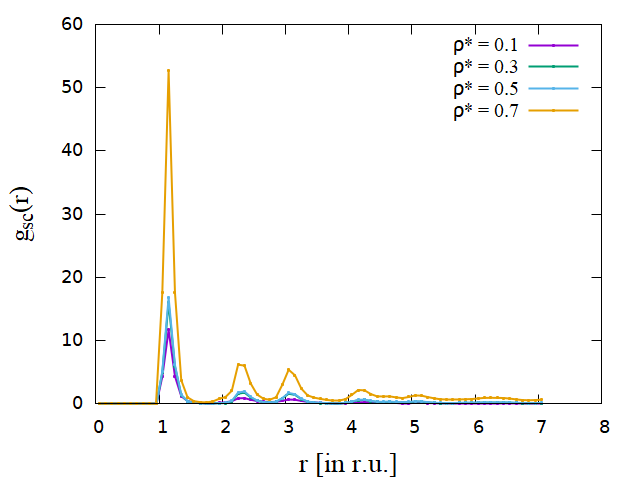}
		\caption{$x = 0.6$}
	\end{subfigure}
	\hspace{2mm}
	\begin{subfigure}{0.48\textwidth}
		\includegraphics[width=0.99\textwidth]{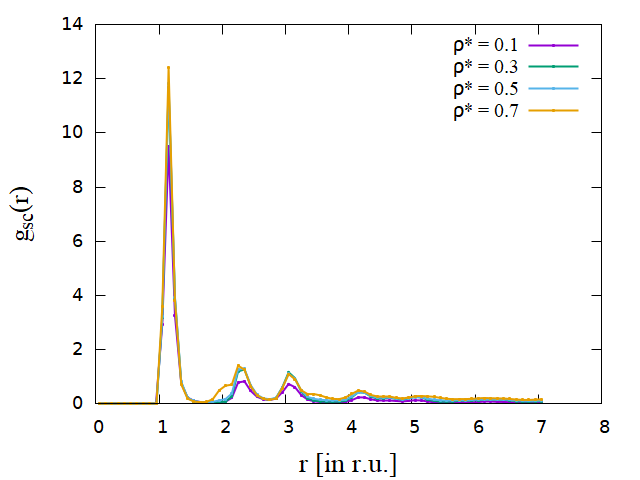}
		\caption{$x = 0.7$}
	\end{subfigure} 
	\caption{(Color online) Radial distribution function between S and C type particles, $g_{sc}(r)$ observed at the specified values of $x$, indicating maximum number of ordered neighbors when $x$ is in the range of 0.5 to 0.6.}
	\label{fig:three}
\end{figure*}

\begin{figure*}
	\centering
	\begin{subfigure}{0.48\linewidth}
		\includegraphics[width=0.99\textwidth]{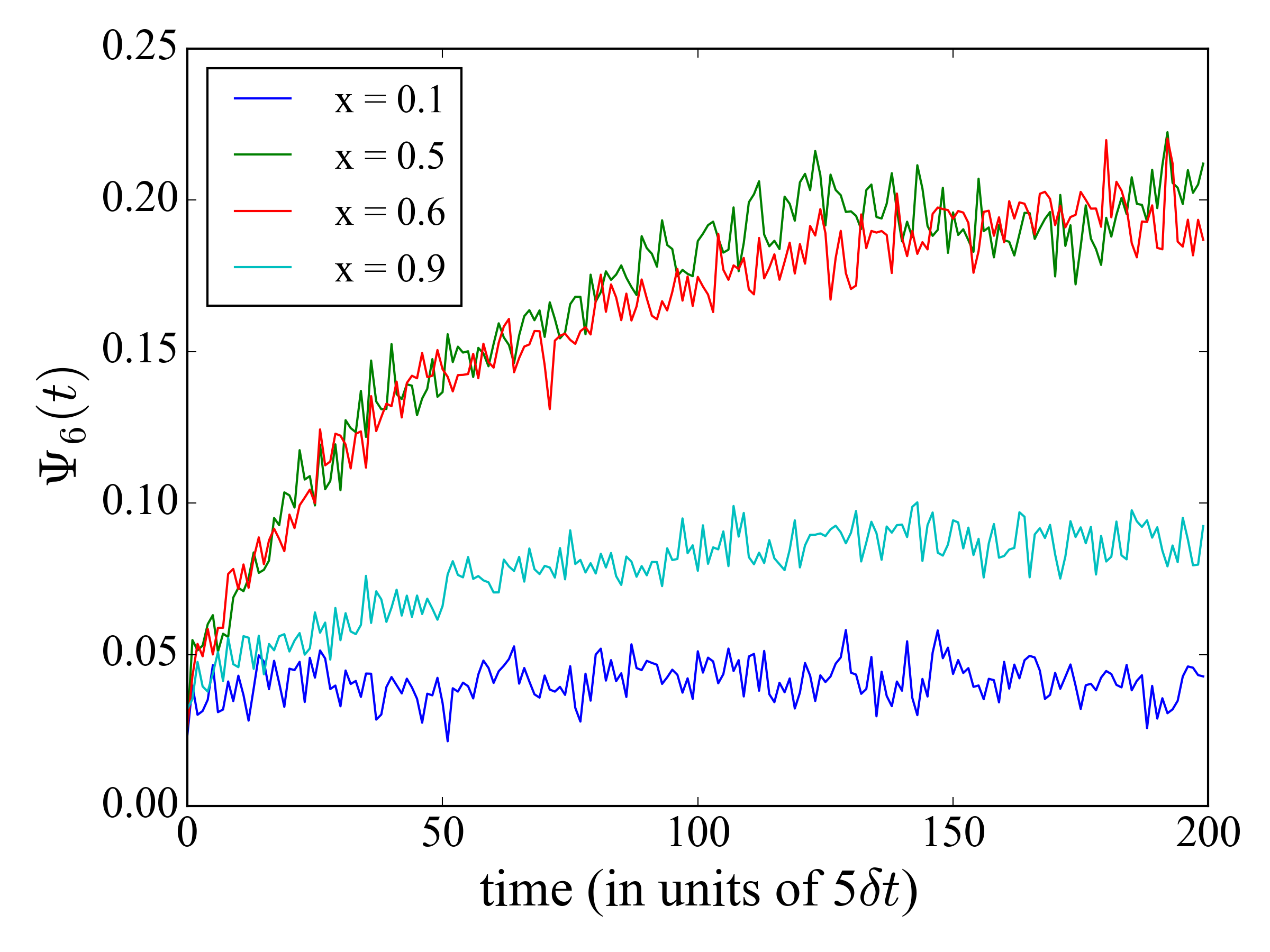}
		\caption{for $\rho^* = 0.1$}
	\end{subfigure}
	\begin{subfigure}{0.48\linewidth}
		\includegraphics[width=0.99\textwidth]{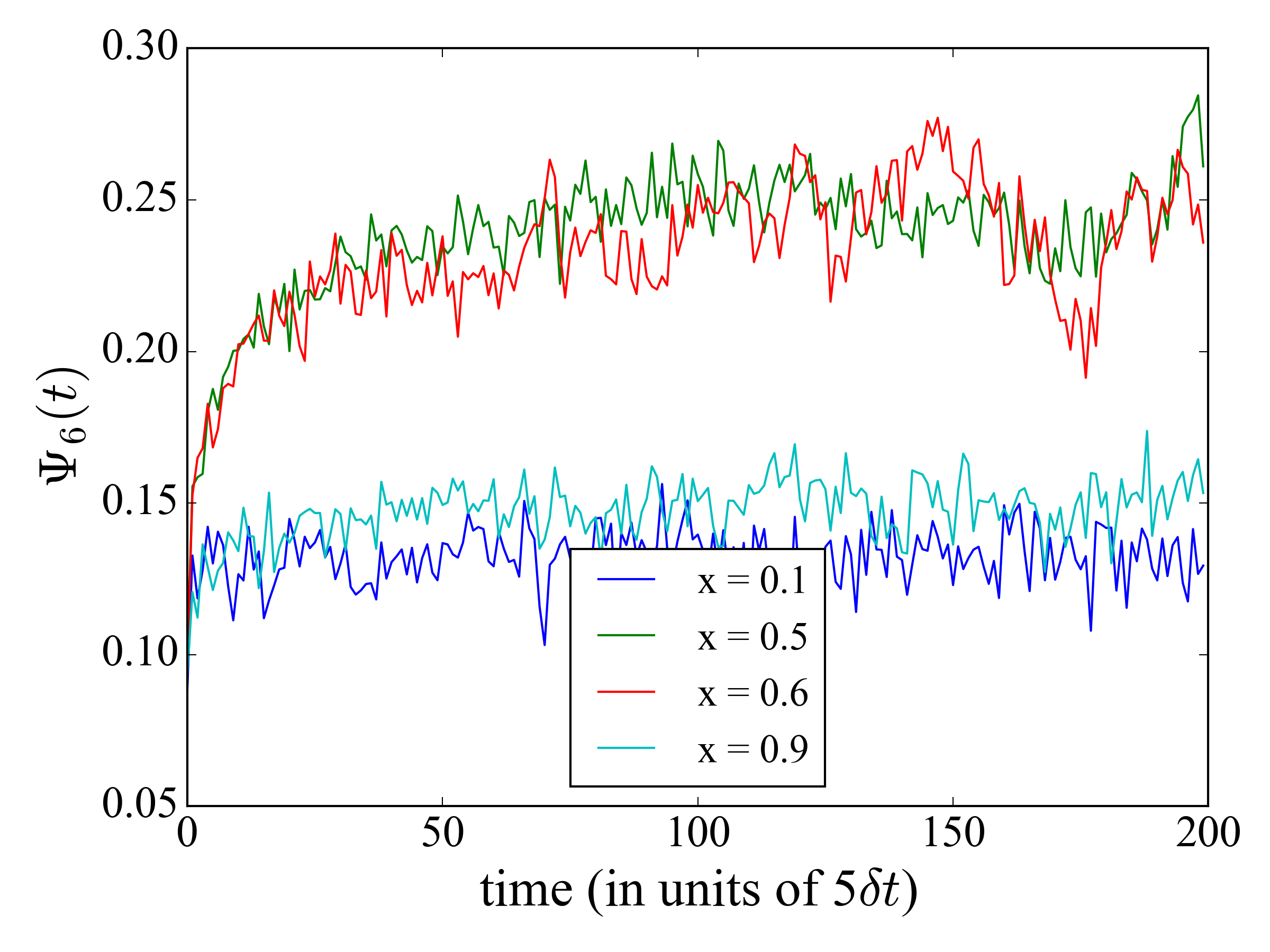}
		\caption{for $\rho^* = 0.3$}
	\end{subfigure} 
	\begin{subfigure}{0.48\linewidth}
		\includegraphics[width=0.99\textwidth]{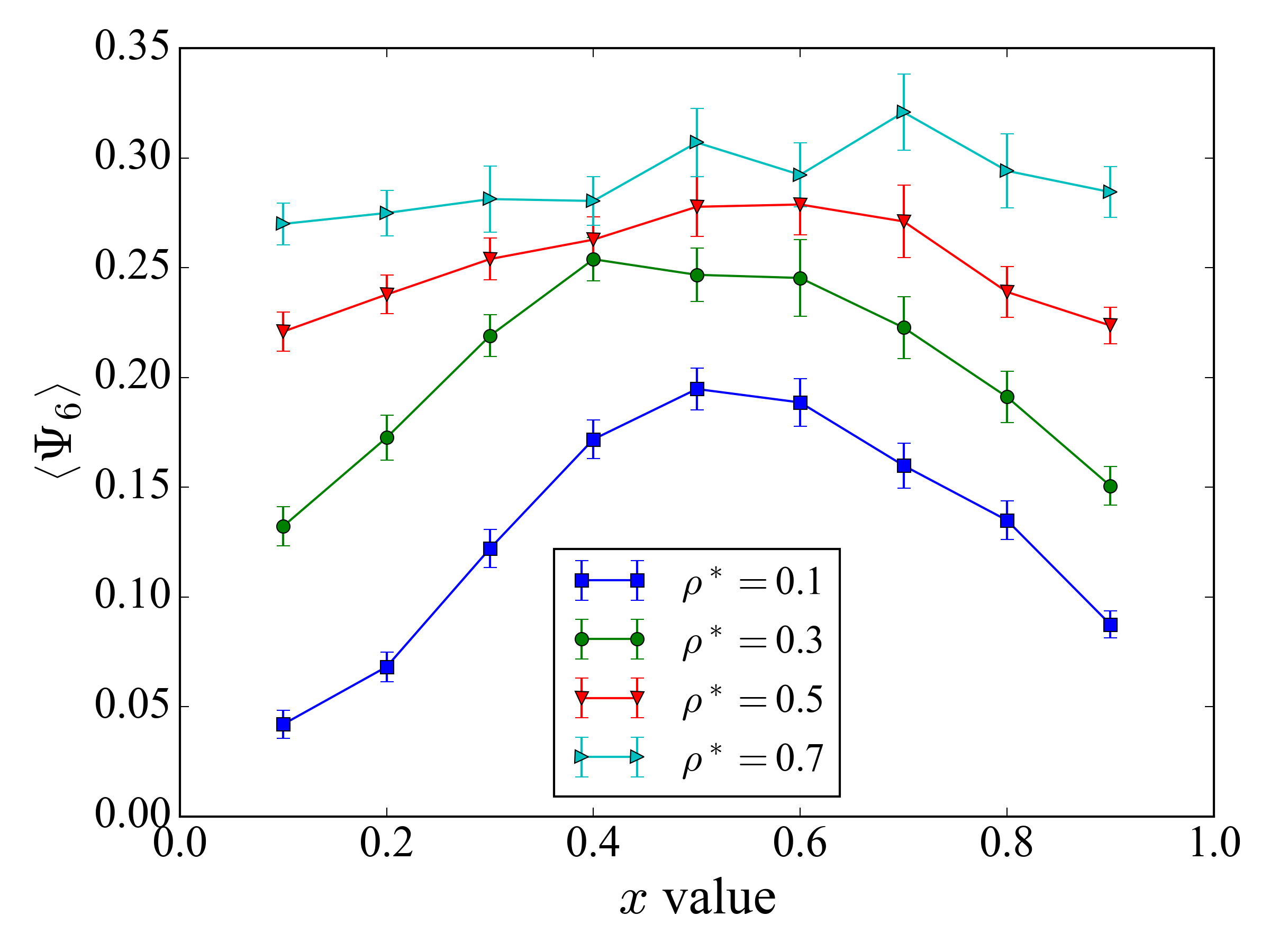}
		\caption{}
	\end{subfigure}
	\caption{(Color online) (a),(b). Time variation of the hexatic order parameter observed at every 5th frame of the simulation trajectory written at a frequency of 1000 time units. (c). Time averaged hexatic order parameter at each $x$. Error bars are obtained from the standard deviation of $\psi_{6}(t)$ during its saturated regime.}
	\label{fig:four}
\end{figure*}

The raft-like network formed between saturated (S-type) and cholesterol (C-type) particles exhibits positional order, as reflected by the radial distribution function $g(r)$ computed between the S- and C-type particles; refer to fig. (\ref{fig:three}). Moreover, the system screenshots in fig. (\ref{fig:one}) indicate the existence of positional order in the range of $x = 0.5 - 0.6$. We also obtained $g(r)$ between the U-S, U-U, and U-C particle types, for $\lbrace x = 0.1, 0.3, 0.6, \rm{and}~ 0.9\rbrace$, which show no significant second shell peaks. See fig. SF4 in the SI for more details. This indicates that no positional order can be seen for the types of particles other than S and C. Meaning, excluding the S-C raft-like complex, no positional order can be seen in the outer fluid-like regime. This is consistent with the earlier experimental \cite{mukai_lipid_2017} and simulation \cite{gomez_actively_2008,sarkar_minimal_2021,smondyrev_structure_1999} observations. 
The hexatic order parameter \cite{nelson_dislocation_1979} is employed to quantify this order,
\begin{equation}
	\psi_{6}^{k} = \frac{1}{n_b}\sum_{j=1}^{n_b}{e^{i6\theta_{jk}}}~.
\end{equation}
Here, $n_b$ is the number of neighbors of $k$th particle within a distance of $1.2\sigma$, and $\theta_{jk}$ is the angle made by $\vec{r}_{jk} = (\vec{r}_{j} - \vec{r}_{k})$ with the $x-$axis. $\psi_{6}^{k}$ is then averaged over all $k$, to get $\psi_{6}(t)$, see fig. \ref{fig:four}(a),(b). $\psi_{6}(t)$ is further time averaged over the last 100 frames of the trajectory, to obtain $\langle \psi_{6} \rangle$, see fig. \ref{fig:four}(c). From the figure, we observe that the extent of the positional order increases with increasing $x$, attaining a maximum at around $x \in \lbrace 0.5 - 0.6 \rbrace$; consistent with the other results of the simulation and the phenomenological calculation.

\section{Conclusion}

Several of the physiological and functional aspects of the cell membrane are directly related to the lateral organization of the lipids in the presence of the relevant protein or cholesterol. The formation of ordered microdomains, or cellular rafts within the membrane eases the cellular interactions with the necessary functional residues. However, there is no complete understanding of the structure and dynamics of such microdomains or lipid rafts. An all-atom atomistic simulation of the relevant system is computationally demanding. The raftlike domains can be understood as a consequence of complex interactions among the constituents. Therefore, here we considered a highly coarse-grained model of the system where each lipid molecule (also cholesterol) is considered spherical in shape and interacts via a modified Lennard-Jones, or a WCA potential. Such a simplified model 
is able to result in the ordered microdomains in the system, composed of the saturated (S-type) and the cholesterol (C-type) particles, that coexist with the fluid-like disordered environment of the unsaturated(U-type) particles at specified mole fractions of C-type. The systems's free energy is found to be minimum at an optimal cholesterol composition of $x \sim 0.6$. The same is obtained from analysis of the phenomenological free energy. These microdomains also acquire maximum hexatic order, around the same value of $x$. These results imply that the system is more stable at around $(x \sim 0.6)$. \\
The seminal work carried out by McConnell in the ternary mixtures of S-,U- and C-type particles suggests to include three particle interactions in such a system. Here in this work, the interaction strength between the S-C types is considered linear in C-type composition $x$, which is a modified pair-wise interaction, not a complete three-particle type interaction though. Despite this, the model is able to predict the optimal C-type composition, which matches with the experimental observation. 

However, a point to note here is that either in a simulation or in an experiment, the exact compositions of cholesterol at which the raft-like ordered microdomains form within a monolayer or a bilayer are highly dependent on the type of lipid molecules and sterols under consideration, and therefore it may be highly subjective to the pertaining interactions therein. But a common feature in all observations is the coexistence of $L_o$ and $L_d$ phases at certain mole fractions of cholesterol, which one can associate with the optimal cholesterol composition(s) for that class of lipids and sterols.

As the lipid molecules are elongated in shape, they can have an orientation, that plays an important role in studying the coexistence of the liquid expanded (LE) and liquid condensed (LC) phases in monolayers. Therefore, the present model can be extended to include the orientation vector for each molecule and define an equation of motion for the same. This might formulate an interesting model to study the liquid ordered-liquid disordered ($L_o - L_d$) transition under the framework of this simplified model, which remains to be the future scope of the work.

\section*{Statements and Declarations}

\begin{itemize}
    \item \textbf{Conflict of interest:} The authors declare no conflict of interest.
    
    \item \textbf{Funding statement:} This research received no specific grant from any funding agency in the public, commercial, or not-for-profit sectors.
    
    \item \textbf{Data availability:} The data that supports the findings of this study is available from the corresponding author upon reasonable request.
    
    \item \textbf{Author contributions:} Both the authors contributed equally to this work and reviewed the manuscript.
\end{itemize}


\bibliography{sn-article}

\end{document}


\maketitle
\section{Variation with the size of the C-type particles}
\label{sec:csizevar}

\begin{figure*}
    \begin{subfigure}{0.48\linewidth}
        \includegraphics[width=0.99\textwidth]{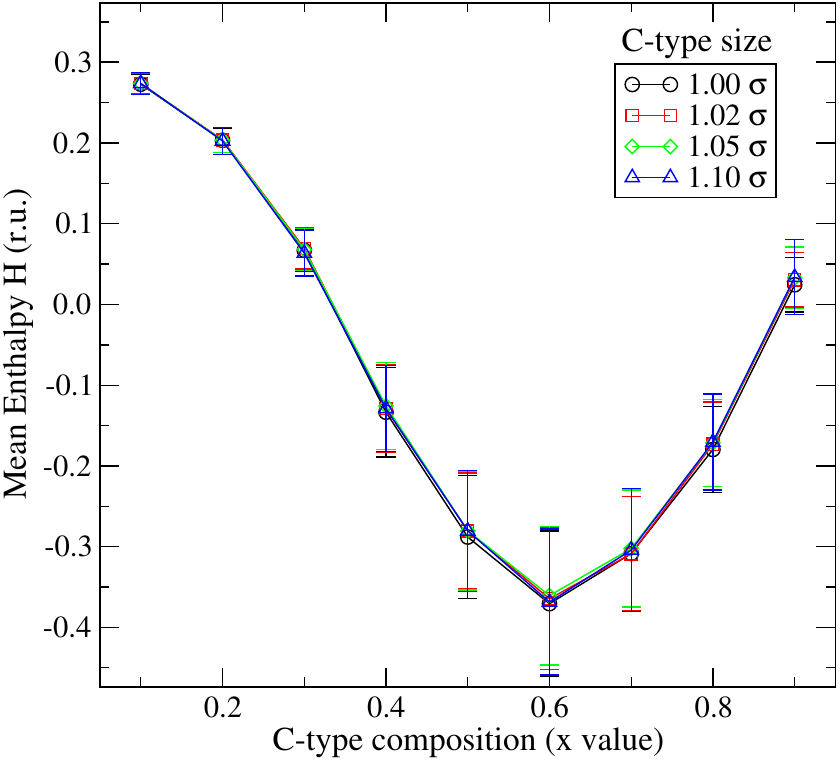}
        \caption{}
    \end{subfigure}
    \hfill
    \begin{subfigure}{0.48\linewidth}
        \includegraphics[width=0.99\textwidth]{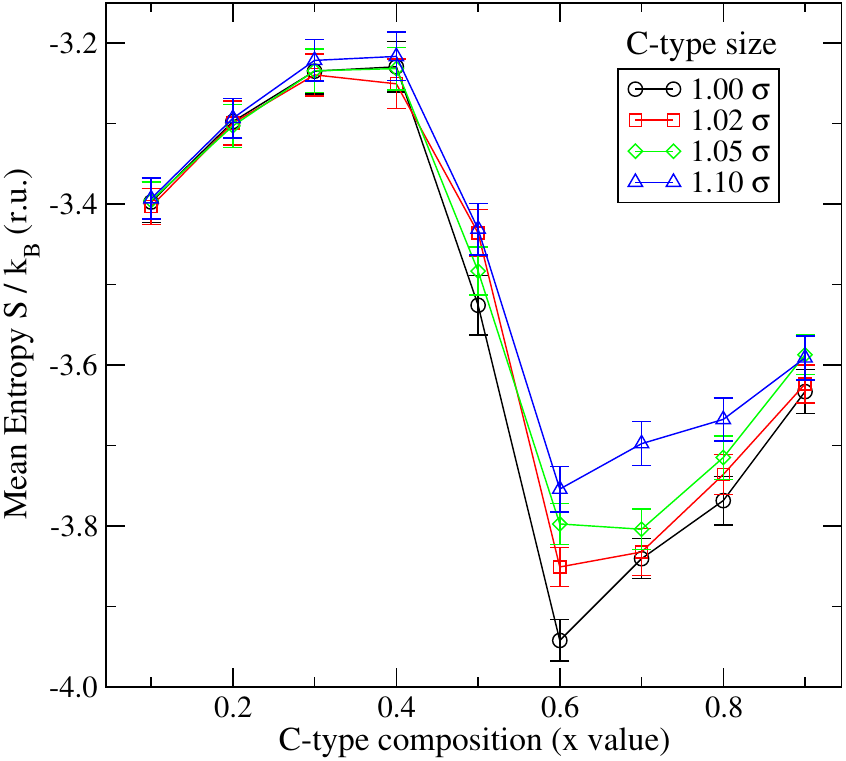}
        \caption{}
    \end{subfigure}
    \caption{(color online) (a) Variation of mean enthalpy and (b) mean entropy as a function of C-type composition for various sizes of the C-type particle, differing by 0\%, 2\%, 5\%, and 10\%.}
    \label{fig:csize}
\end{figure*}

We have conducted the simulations by systematically varying the size of the C-type particle that differs by {0\%, 2\%, 5\%, and 10\%.} from the other types. Respective thermodynamic parameters are calculated, and the behavior of enthalpy and entropy as a function of the composition variable $x$ is as depicted in fig. \ref{fig:csize}.

From the above figure, we notice that all these cases show almost no difference in enthalpy. However, the entropy picks up noticeable differences at $x^*$ and its neighboring composition values, with an increase in the size of the C-type. So, further higher values of C-type size could lead to increased entropy values, which may favor mixing rather than the complex formation and the associated positional ordering. Therefore, a judicious size difference of 1\% is assumed initially to create a minimal possible difference between the lipid molecules and cholesterol sizes.

\section{Variation with the type of thermostat}
\label{sec:thermovar}

\begin{figure*}
    \includegraphics[width=1.0\linewidth]{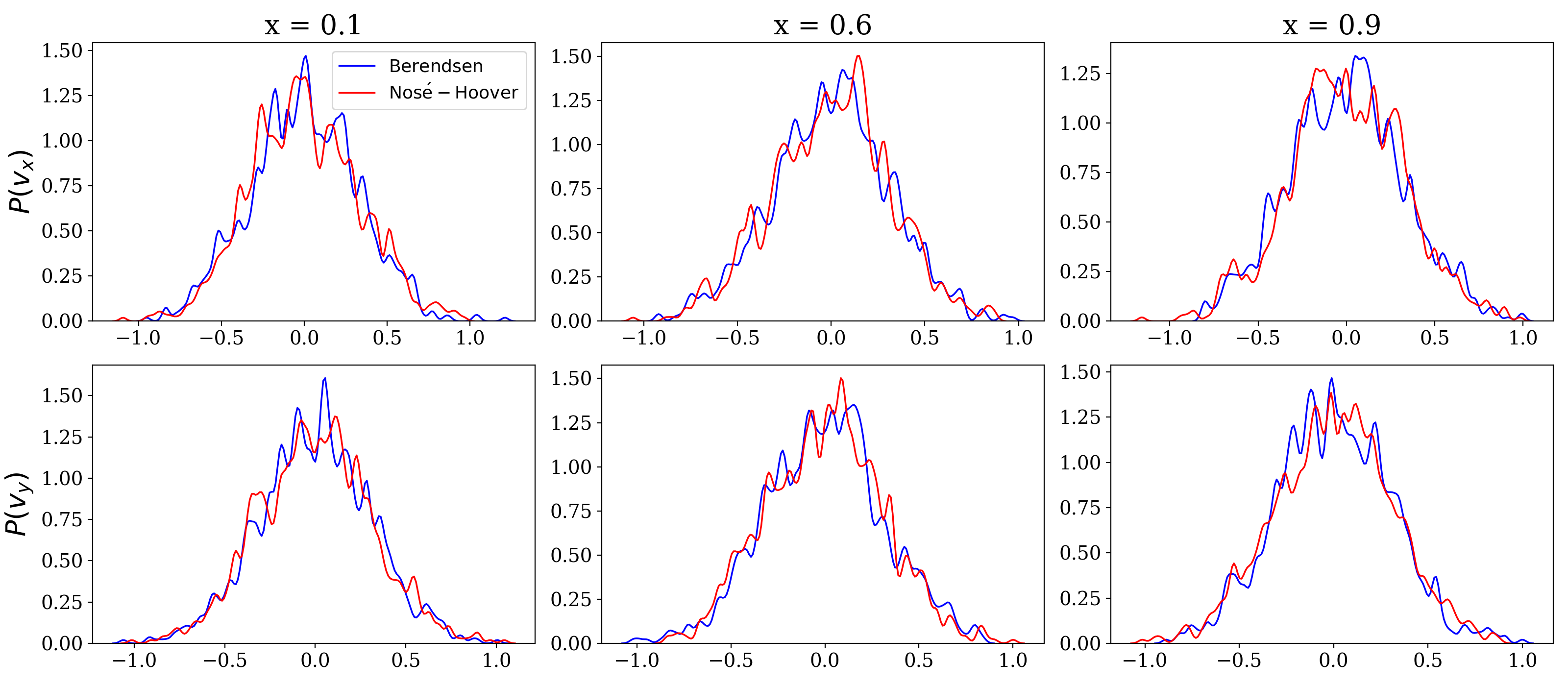}
    \caption{(color online) Normalized density distributions of the x- and y-components of velocity, $P(v_x)$, and $P(v_y)$, obtained at the specified $x$ values, pertaining to the Nos\'e-Hoover and the Berendsen thermostats. These are obtained for the density of $\rho^{*} = 0.3$, and a 2\% difference for C-type particle size.}
    \label{fig:hist_vxvy}
\end{figure*}

\begin{figure*}
    \centering
    \begin{subfigure}{0.48\linewidth}
        \includegraphics[width=0.99\textwidth]{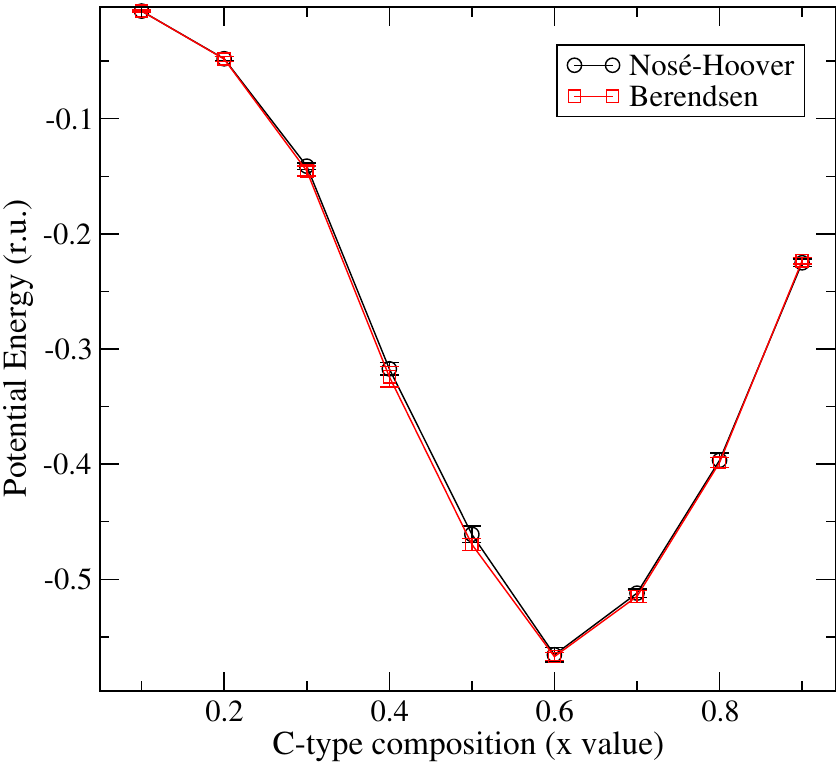}
        \caption{}
    \end{subfigure}
    \hfill
    \begin{subfigure}{0.48\linewidth}
        \includegraphics[width=0.99\textwidth]{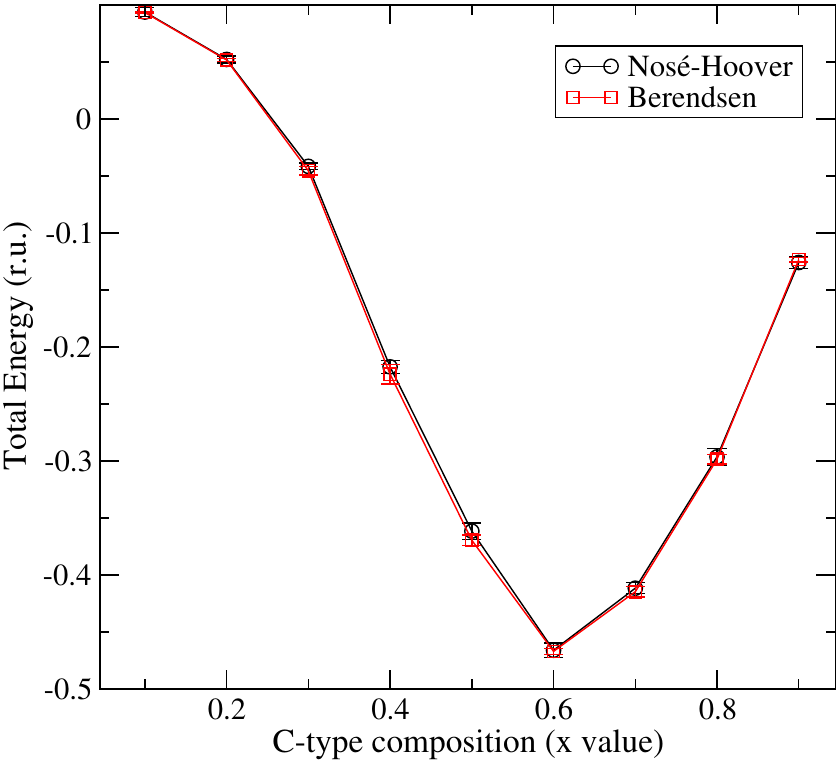}
        \caption{}
    \end{subfigure} \\[2mm]
    \begin{subfigure}{0.48\linewidth}
        \includegraphics[width=0.99\textwidth]{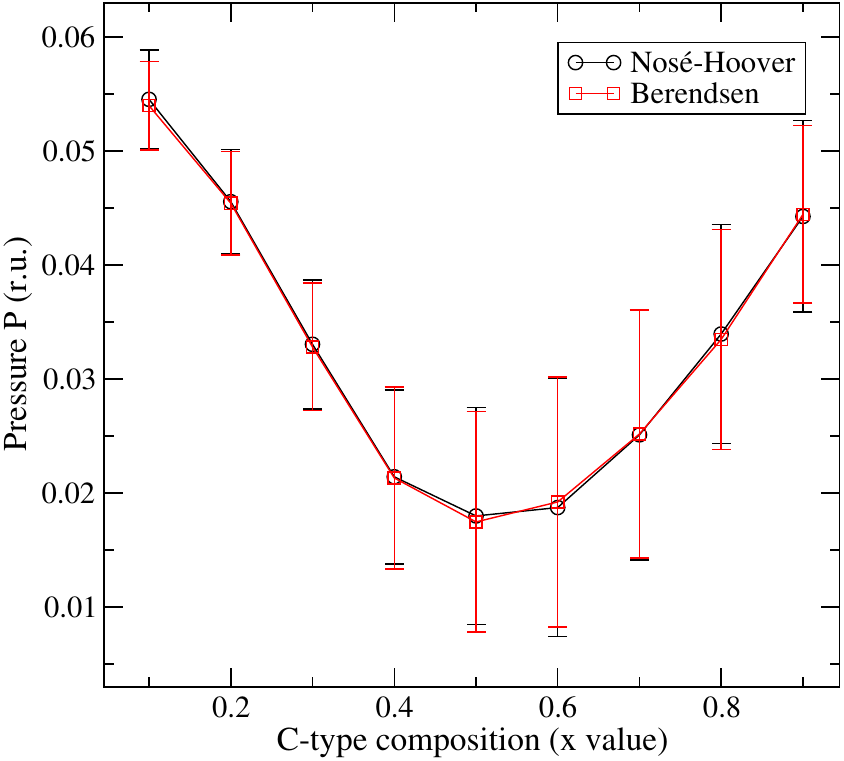}
        \caption{}
    \end{subfigure}
    \hfill
    \begin{subfigure}{0.48\linewidth}
        \includegraphics[width=0.99\textwidth]{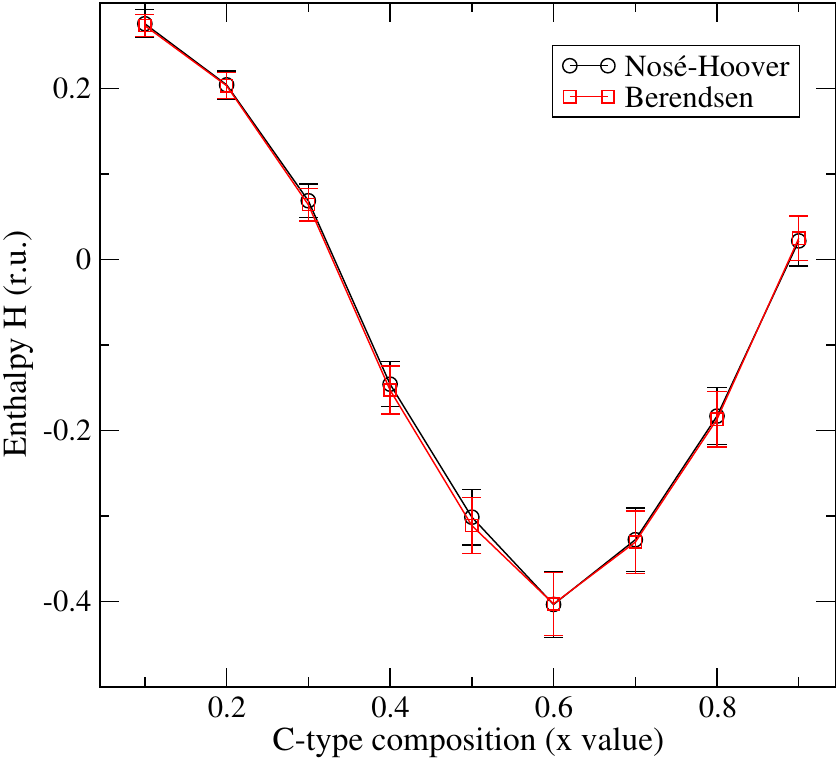}
        \caption{}
    \end{subfigure} \\[2mm]
    \begin{subfigure}{0.48\linewidth}
        \includegraphics[width=0.99\textwidth]{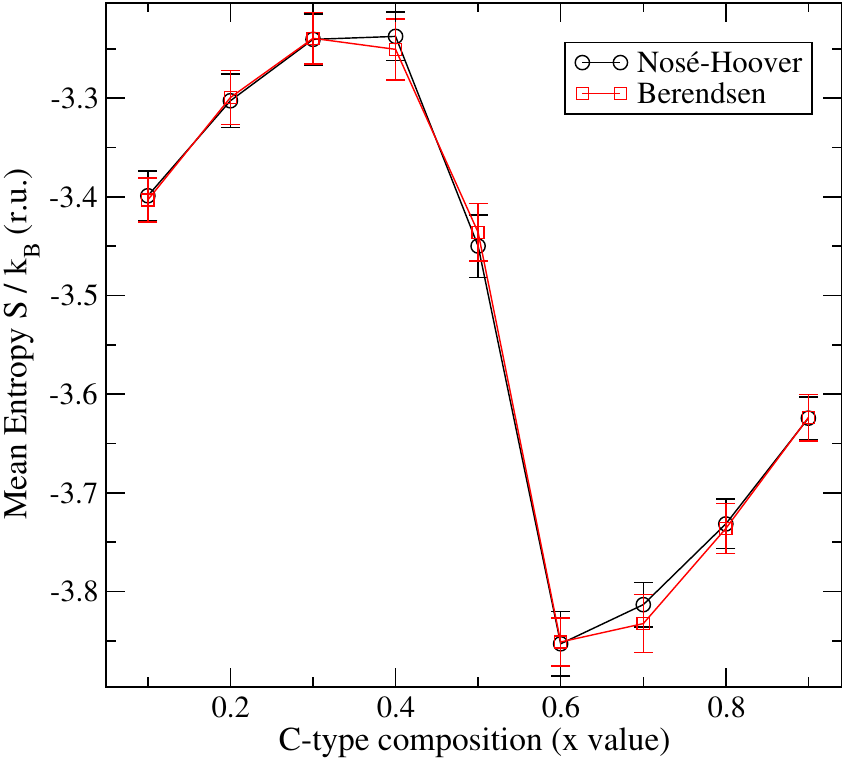}
        \caption{}
    \end{subfigure}
    \caption{(color online) Variation of time-averaged (a) potential energy, (b) total energy, (c) pressure, (d) enthalpy, and (e) entropy with the C-type composition, obtained using both the Nos\'e-Hoover and Berendsen thermostats. These simulations are done at a density of $\rho^{*} = 0.3$, and a 2\% difference for C-type size.}
    \label{fig:thermo}
\end{figure*}

We have performed the simulations using the Nos\'e-Hoover and the Berendsen thermostats, and the results of the same are shown in fig. \ref{fig:hist_vxvy} and \ref{fig:thermo}. 
Figure \ref{fig:hist_vxvy} shows the normalized density distributions of the velocity components obtained from one of the system configurations. No significant difference is found between both the thermostats.

Figure \ref{fig:thermo} depicts the behavior of thermodynamic variables of the system as a function of the C-type composition for both the thermostats. \\
The Berendsen thermostat involves a velocity rescaling mechanism with the associated coupling parameter $\tau$ of the system to the reservoir, whereas the Nos\'e-Hoover thermostat extends the Hamiltonian to include the additional degrees of freedom pertaining to the reservoir. Therefore, the Berendsen thermostat relaxes the system faster to the targeted temperature than the Nos\'e-Hoover thermostat. 

\section{Radial distribution function between distinct particle types}
\label{sec:gofr}

\begin{figure*}
    \includegraphics[width=1.0\linewidth]{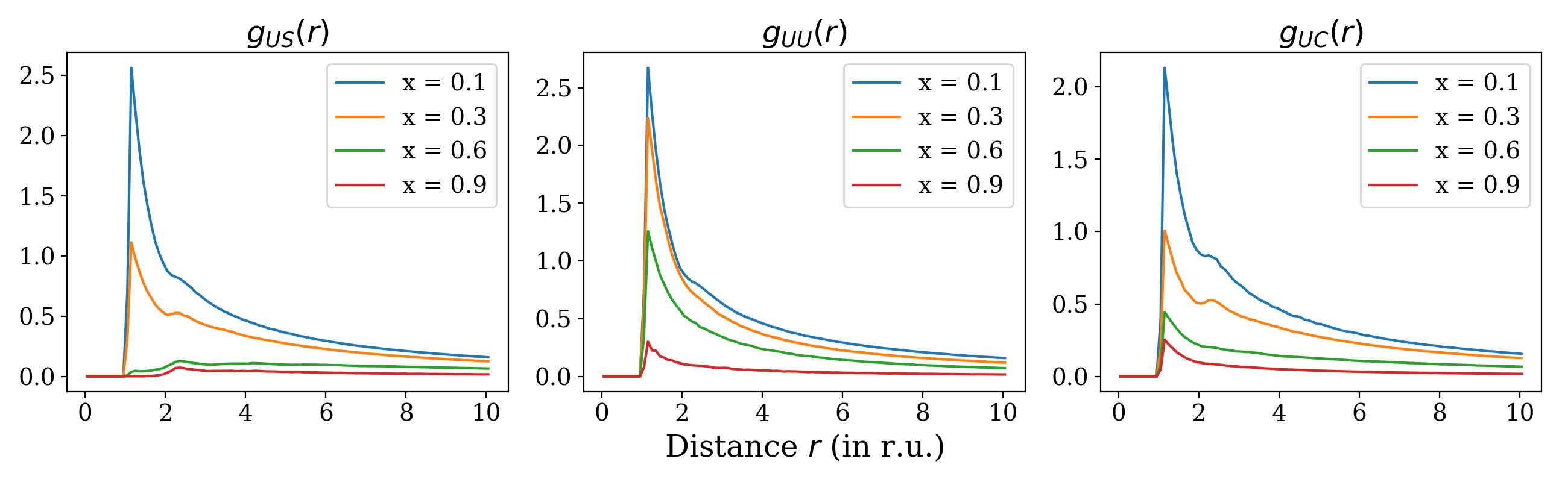}
    \caption{(color online) Radial distribution functions, for U-S, U-U, and U-C type pairs of particles in the system, obtained for specified values of ${x = 0.1, 0.3, 0.6, \rm{and}~  0.9}$.}
    \label{fig:gofr_si}
\end{figure*}

The radial distribution function, obtained between the various particle types and averaged over a trajectory of system configurations carries the signature of the dynamics of the constituents. Here, in fig. \ref{fig:gofr_si}, we have depicted the radial distribution function obtained between the U-S, U-U, and U-C particle types for $\lbrace x = 0.1, 0.3, 0.6, \rm{and}~ 0.9\rbrace$. The interaction potential defined in the methodology section of the manuscript reveals that U-type particles only had ‘repulsive’ interactions with all types of particles. The same is evident in the behavior of $g_{US}(r), g_{UU}(r)$, and $g_{UC}(r)$. They show a decaying type of feature with distance $r$. This indicates that the ordered arrangement of particles, such as noticed in $g_{SC}(r)$ (in the main manuscript), is not present among the other type of particle pairs. As the U-type are mostly present outside the ordered domain, we can conclude that no positional order could be seen in the fluid-like region outside the ordered domains. This is consistent with the earlier experimental \cite{mukai_lipid_2017} and simulation \cite{gomez_actively_2008,sarkar_minimal_2021,smondyrev_structure_1999} observations.

\bibliographystyle{ieeetr}
\bibliography{sn-article}